\documentstyle[12pt]{article}

\begin{document}



\newcommand {\beq}{\begin{equation}}
\newcommand {\eeq}{\end{equation}}
\newcommand {\beqa}{\begin{eqnarray}}
\newcommand {\eeqa}{\end{eqnarray}}
\newcommand {\beqan}{\begin{eqnarray*}}
\newcommand {\eeqan}{\end{eqnarray*}}
\newcommand {\n}{\nonumber \\}
\newcommand {\eq}[1]{eq.~(\ref{#1})}
\newcommand {\eqs}[1]{eqs.~(\ref{#1})}

\newcommand {\del}{\partial}
\newcommand {\eV}{\:{\rm eV}}
\newcommand {\keV}{\:{\rm keV}}
\newcommand {\MeV}{\:{\rm MeV}}
\newcommand {\GeV}{\:{\rm GeV}}
\newcommand {\TeV}{\:{\rm TeV}}

\def\e{\epsilon}
\begin{titlepage}
 \renewcommand{\thefootnote}{\fnsymbol{footnote}}
    \begin{normalsize}
     \begin{flushright}
                KEK-TH-617 \\
                March 1999 \\
      \end{flushright}
    \end{normalsize}
    \begin{Large}
       \vspace{1cm}
       \begin{center}
         {\bf Space-Time and Matter \\ in  IIB Matrix Model \\
          -  gauge symmetry and diffeomorphism  -
         } \\
       \end{center}
    \end{Large}

  \vspace{10mm}
\begin{center}
\begin{Large}
           S. Iso\footnote
           { e-mail : satoshi.iso@kek.jp} and
           H. Kawai\footnote{e-mail : hikaru.kawai@kek.jp, \ 
            Address after April 1999, Kyoto University}

\end{Large}
         {\it High Energy Accelerator Research Organization} \\
             {\it KEK} 
              {\it Oho 1-1, Tsukuba, Ibaraki, Japan } \\      
\vspace{15mm}

\end{center}
\begin{abstract}
\noindent 
We pursue the study of the type IIB matrix model as a constructive definition
of  superstring.
In this paper, we justify the interpretation of  
space-time as distribution of  eigenvalues  of  the matrices 
by showing that some low energy excitations  indeed propagate in it.
In particular, we show that if the distribution  consists of 
small clusters of size $n$,  low energy  theory acquires local
$SU(n)$ gauge symmetry and  a plaquette action for the 
associated gauge boson  is induced, 
in addition to a gauge invariant kinetic term for a massless
fermion in the adjoint representation of the $SU(n)$.
We finally argue a possible identification 
of the  diffeomorphism 
symmetry with  permutation group acting on the set of eigenvalues,
and  show that the general covariance is realized  in  the low
energy effective theory even though
we do not have a manifest general covariance in the IIB matrix model action.
\end{abstract}

\end{titlepage}
\vfil\eject

\newpage
\section{Introduction}
\setcounter{equation}{0}
Several proposals have been given as constructive definitions 
of  superstring 
\cite{IKKT,BFSS,DVV,periwal,yoneya,Polch,sugawara,Itoyama,hetero,IIBvariant}. 
Type IIB matrix model \cite{IKKT,FKKT,AIKKT},
a large $N$ reduced model of maximally supersymmetric Yang-Mills theory,
is one of those proposals.
It is defined by the following action:
\beq
S  =  -{1\over g^2}Tr({1\over 4}[A_{\mu},A_{\nu}][A^{\mu},A^{\nu}]
+{1\over 2}\bar{\psi}\Gamma ^{\mu}[A_{\mu},\psi ]) ,
\label{action}
\eeq
where
$A_{\mu}$ and $\psi$ are $N \times N$ Hermitian matrices,
$\mu=1,,,10$ and  $\psi$ is a ten dimensional Majorana-Weyl spinor
field.
It is formulated in a manifestly covariant way,  which 
is suitable for studying 
 nonperturbative issues of superstring theory.
Also  since it is a simple  model of an ensemble of  zero dimensional matrices,
it is  particularly appropriate for numerical simulations.
In fact we can in principle predict  dimensionality of space-time,
low energy gauge group and  matter contents by solving this model.
In this paper, we pursue this line of analysis and report some results
on  possible structures of  the low energy effective theory  and 
the origin of 
local space-time gauge symmetry and   diffeomorphism invariance. 
These analysis gives justification of our interpretation of space-time
as  distributed eigenvalues of the matrices $A_{\mu}$.
\par
We first list several important properties of the IIB matrix model.
This model can be  regarded as a large $N$ reduced model of ten
dimensional ${\cal N}=1$
supersymmetric $SU(N)$ Yang-Mills theory. 
It was shown  \cite{RM} that a large $N$ gauge theory 
can be equivalently described by its reduce model, namely  a model
defined on a single point. In this reduction procedure  space-time 
translation is represented in the color $SU(N)$ space and eigenvalues
of matrices are interpreted as momenta of fields.
As a constructive definition of superstring, on the other hand, 
we will see  that we need to interpret eigenvalues of matrices
as coordinates of space-time points, which interpretation 
is T-dual to the above. 
\par
Since our IIB matrix model is defined on a single point, 
the commutator of the  supersymmetry
\beq
\cases{
\delta^{(1)} A_\mu &  $= i \bar{\epsilon_1}\Gamma_\mu \psi $\cr
\delta^{(1)} \psi & $=\frac{i}{2}\Gamma^{\mu\nu} [A^\mu, A^\nu] {\epsilon_1}$
},  \label{Ssym1}
\eeq
vanishes up to a field dependent 
gauge transformation and we can no longer interpret
this supersymmetry as  space-time supersymmetry in the original sense.
However, after the reduction, we acquire  an extra bosonic symmetry
\begin{equation}
\delta A_{\mu} = c_{\mu}  {\bf 1},
\label{Tsym}
\end{equation}
whose transformation is 
proportional to a unit matrix ${\bf 1}$
and an extra supersymmetry 
\beq
\cases{
\delta^{(2)} A_\mu & = 0  \cr
\delta^{(2)} \psi & $=\epsilon_2$
}.
\label{Ssym2}
\eeq
Combinations of  these two supersymmetries (\ref{Ssym1}) and (\ref{Ssym2})
\begin{equation}
\tilde{Q}^{(1)} = Q^{(1)} + Q^{(2)}, \ 
\tilde{Q}^{(2)} = i( Q^{(1)} - Q^{(2)}),
\end{equation}
 satisfy commutation relations
\begin{equation}
[\bar{\epsilon_1} \tilde{Q^{(i)}}, \bar{\epsilon_2} \tilde{Q^{(j)}}]
= -2 \bar{\epsilon_1} \gamma_{\mu} \epsilon_2 p^{\mu} \delta^{(ij)}
\end{equation}
where $p^{\mu}$ is the generator of the translation (\ref{Tsym})
and $i,j= 1, 2$.
Therefore, if we  interpret  eigenvalues of the  matrices $A_{\mu}$
as our new space-time coordinates, 
the above symmetries
can be regarded as  ten-dimensional ${\cal N}=2$ space-time
supersymmetries.
Since the maximal space-time supersymmetry  guarantees  the existence of
graviton,  it supports the conjecture  that the IIB matrix model
is a constructive definition of superstring.
This is one of the major reasons to interpret the eigenvalues of 
$A_{\mu}$ as  coordinates of the  newly emerged space-time.
\par
The second important and confusing  property 
is that the model has the same 
action as  the low energy effective action of D-instantons 
\cite{Witten}.
We should emphasize here  differences between these two theories
since we are led to different interpretations of space-time.
From the effective theory point of view,  the eigenvalues represent
the coordinates of D-instantons in the ten
dimensional  bulk space-time  which we have assumed a priori
from the beginning of constructing the effective action. 
On the other hand, 
from the  constructive point of view we cannot assume such a 
bulk space-time itself in which matrices live, 
since not only fields but also the
space-time should be  dynamically generated as a result of dynamics of
the matrices and  should be constructed only from the  matrices.
The most natural interpretation is that the space-time consists of
$N$ discretized  points and  the eigenvalues represent their
space-time coordinates.
One of  the main purposes  of this paper is to 
confirm this 
interpretation of the  space-time 
and make sure that low energy excitations propagate in this space-time.
\par
Final important property is that the type IIB matrix model has no
free parameters. The coupling constant $g$ can be always absorbed by 
field redefinitions:
\begin{eqnarray}
\cases{
A_{\mu} \rightarrow g^{1/2} A_{\mu} \cr
\psi \rightarrow g^{3/4} \psi.
}
\end{eqnarray}
This is reminiscent of  string theory where a  shift of the string coupling
constant is always absorbed to that of the  dilaton vacuum expectation
value (vev).
In the previous analysis of the Schwinger-Dyson equation 
of IIB matrix model \cite{FKKT}, 
we have introduced an infrared cut-off $\epsilon$,  which gives  string
coupling constant $g_{st} = 1/N\epsilon^2$. But through more careful analysis
of  dynamics of the eigenvalues \cite{AIKKT},  we have shown that there is 
no such  infrared divergences associated with  infinitely 
separated eigenvalues and the infrared cutoff $\epsilon$ we have
introduced
by hand can be 
determined dynamically in terms of $N$ and $g$. 
Roughly speaking, as we will study in the
present paper, the
density of the eigenvalues represents  dilaton vev and gives the
string coupling constant. More solid relation requires  
knowledge of  double scaling limit and we hope to report 
in a separate paper. 
\par
There are several reasons we believe that IIB matrix model is a 
constructive definition of the type IIB superstring. 
First
this action can be related to the Green-Schwarz action of 
superstring\cite {GS}
by using  the semiclassical correspondence in the large $N$ limit:
\beqa
-i[\;,\;] &\rightarrow&  \{\;,\;\}, \n
Tr &\rightarrow&  \int {d^2 \sigma }\sqrt{\hat{g}} .
\label{correspondence}
\eeqa
In fact eq.(\ref{action}) reduces to the Green-Schwarz action
in the Schild gauge\cite{Schild}:
\beq
S_{\rm Schild}=\int d^2\sigma [\sqrt{\hat{g}}\alpha(
\frac{1}{4}\{X^{\mu},X^{\nu}\}^2
-\frac{i}{2}\bar{\psi}\Gamma^{\mu}\{X^{\mu},\psi\})
+\beta \sqrt{\hat{g}}].
\label{SSchild}
\eeq
Through this correspondence,  the eigenvalues of $A_{\mu}$ matrices are
identified with  the space-time coordinates $X_{\mu}(\sigma )$ of string
world sheet.
This is consistent with our interpretation that the eigenvalues
of the matrices represent  space-time coordinates.
\par
The correspondence can go farther 
beyond the above identification of the  model with  a matrix regularization 
of the first quantized superstring.
Namely, we can describe arbitrary  number of interacting D-strings 
and anti-D-strings as blocks of matrices, each of which corresponds
to matrix regularization of a string.
Off-diagonal blocks induce interactions between  these strings
\cite{IKKT,effective}.
Thus it must be clear that the IIB matrix model is definitely not the first
quantized theory of a D-string but the full second quantized theory.
\par
It has also been  shown \cite{FKKT} that  Wilson loops satisfy  the
string field equations of motion for the type IIB superstring in the 
light cone gauge, which is the second evidence for the conjecture
that the IIB matrix model is a constructive definition of strings.
We consider the following regularized Wilson loop \cite{FKKT,hamada}:
\beq
w(C)  =  Tr[\prod_{n=1}^M exp\{ i\epsilon (k_{n}^{\mu} A_{\mu}
+\bar{\lambda}_{n} \psi ) \} ] .
\label{Wilsonloop}
\eeq
Here $k_n^{\mu}$ are momentum densities distributed along a loop $C$, and
we have also introduced  fermionic sources
$\lambda _{n}$ .
$\epsilon$ in the argument of the exponential
is an infrared cutoff  associated to  space-time extension.
In the large $N$ limit, 
$\epsilon$ should  go to $0$ so as to satisfy the double scaling limit.
There is still some subtly how to take 
this double scaling limit in which  we get an interacting 
string theory. 
\par
Considered as a matrix regularization of the Green Schwartz IIB superstring,
IIB matrix model describes interacting D-strings. On the other hand
in the analysis of  the  Wilson loops IIB matrix model describes
joining and splitting interactions of fundamental IIB superstrings
as Wilson loops.
From these considerations, it is plausible to conclude
that if we can take the 
correct double scaling limit,  IIB matrix model becomes  a constructive
definition of the type IIB superstring. Furthermore we believe that
all string theories are connected by duality transformations, and once
we  construct a nonperturbative  definition of any one of them, 
 we can 
describe vacua of any other strings, in particular, the true vacuum in 
which we live. 
\par
Once we are convinced that  IIB matrix model is a constructive
definition of superstring,  we can simply go on to do numerical
calculations \cite{numerical}. 
But before doing so  we will  clarify in this paper
how we can read  the space-time geometry from matrix configurations
and also propose a
 scenario of obtaining local space-time gauge symmetry
and  diffeomorphism symmetry. 
It is  amusing that these fundamental symmetries can arise from
a very simple matrix model defined on a single point.
\par
Dynamics of eigenvalues, that is, dynamical generation 
of  space-time  was first discussed in our 
previous paper  \cite{AIKKT}. 
An effective action of  eigenvalues can be 
obtained by integrating all the off-diagonal bosonic and fermionic components
and then diagonal fermionic coordinates (which we call fermion zeromodes). 
If we quench the bosonic diagonal components
$x_{\mu}^{i}$ ($i=1...N$)
\footnote{
In this paper we consider matrix configurations near 
simultaneously diagonalizable ones.
Then the eigenvalues and the diagonal components of matrices coincide
and we  obtain the same effective action for them
at least up to one-loop perturbation in off-diagonal components.
We expect that the classical space-time picture is valid only in such a case.
For more general configurations, the
  classical space-time picture is  broken and 
non-commutativity of space-time will 
become important \cite{Connes}. We leave it for future analysis. 
} 
and neglect the 
fermion zeromodes $\xi^{i}$, the effective action for $x_{\mu}^{i}$ 
coincides with that of 
the $D=4$  $ {\cal N}=4$ supersymmetric Yang Mills theory and vanishes
respecting  stability of  supersymmetric moduli.
Inclusions of fermion zeromodes and also of non-planar contributions
lift the degeneracy and we can obtain a nontrivial effective action
for space-time dynamics. 
In our previous  paper \cite{AIKKT} we estimated this effective action 
by perturbation at one loop, which is valid when all eigenvalues
are far from one another  $|{\bf x}_i-{\bf x}_j| >> \sqrt{g}$. 
Of course this one-loop effective action is not sufficient to 
determine the full space-time structure,  but we expect  
that it captures
some of the essential points for the  formation of the space-time.
One of  important properties of the effective action is that
as a result of  grassmannian integration of the fermion zeromodes,
 space-time points make a network connected locally by bond interactions.
This becomes important when we extract diffeomorphism symmetry 
from our matrix model. This is discussed in section 4.
\par
The organization of the paper is as follows. In section 2, we 
briefly review our previous results on  the dynamics of  the space-time
and show how a  network picture of the space-time arises.
Here is an analogy with the dynamical triangulation approach to 
quantum gravity. 
In section 3 we derive   a low energy effective theory 
in some particular eigenvalue distributions consisting
of small clusters of size $n$. The effective theory acquires
local $SU(n)$ gauge symmetry  and we show that the associated
gauge field indeed propagate in the  distribution.
This supports our interpretation of  the space-time. Furthermore  we show 
that a massless fermion field in the  adjoint representation appears
with  a gauge invariant kinetic term. 
In this way,   low energy effective theory  is formulated as 
a lattice gauge theory on a dynamically generated random lattice.
In section 4, we study the  origin of gravity and diffeomorphism
symmetry  in the  space-time.
We show that  invariance under  permutations of the eigenvalues leads to
the  diffeomorphism invariance.  Background metric is  shown to 
be encoded 
in the density correlation of the eigenvalues.
Section 5 is devoted to conclusions and discussions. 
\section{Dynamics of Eigenvalues and Space-time Generation}
\setcounter{equation}{0}
In this section we briefly review our previous analysis
\cite{AIKKT} on  the
dynamics of  the eigenvalues to  clarify some of important properties. 
Let us consider  expansion around the most generic classical
moduli where the gauge group $SU(N)$ is  broken down to
$U(1)^{N-1}$.
Then  diagonal elements of $A_{\mu}$ and $\psi$ appear
as  zeromodes while  the off-diagonal elements become massive.
We may hence integrate out the massive modes first and obtain an effective
action for the diagonal elements.

We thus decompose $A_{\mu}$ into diagonal part $X_{\mu}$
and off-diagonal part $\tilde{A}_{\mu}$.
We also decompose $\psi$ into $\xi$ and $\tilde{\psi}$:
\beqa
A_{\mu}&=& X_{\mu} + \tilde{A}_{\mu}; \;\;
X_{\mu}= \left( \begin{array}{llll}
                 x_{\mu}^1 &&& \\
                 &x_{\mu}^2 && \\
                 && \ddots &   \\
                  &&& x_{\mu}^N
         \end{array} \right), \n
\psi &=& \xi + \tilde{\psi}; \;\;
\xi = \left( \begin{array}{llll}
                 \xi^1 &&& \\
                 &\xi^2 && \\
                 && \ddots &   \\
                  &&& \xi^N
         \end{array} \right),
\eeqa
where $x^i_\mu$ and $\xi^i_\alpha$ satisfy the constraints 
$\sum_{i=1}^N x^i_\mu=0$ and $\sum_{i=1}^N \xi^i_\alpha =0$, respectively,
since the gauge group is $SU(N)$.
We then integrate out the off-diagonal parts $\tilde{ A_{\mu}}$
and $\tilde{\psi}$ and obtain the
effective action for supercoordinates of space-time $S_{\rm eff}[X,\xi]$.
The effective action for the space-time coordinates $S_{\rm eff}[X]$
can be obtained by further integrating out $\xi$:
\beqa
\int dAd\psi e^{-S[A,\psi]}
&=& \int dXd\xi e^{-S_{\rm eff}[X,\xi]} \nonumber \\
&=& \int dX e^{-S_{\rm eff}[X]},
\eeqa
where $dX$ and $d\xi$ stand for $\prod_{i=1}^{N-1}\prod_{\mu=0}^{9} dx^i_\mu$
and $\prod_{i=1}^{N-1}\prod_{\alpha=1}^{16} d\xi^i_\alpha$, respectively.
Perturbative  expansion in $g^2$ around diagonal backgrounds 
$(X_{\mu}, \xi)$ is valid
when all of the diagonal elements are widely separated from one another:
$|x^i -x^j| >> g^{1/2}$. At one loop, it is easy to integrate over the
off-diagonal components.
\par
After adding a gauge fixing and the Fadeev-Popov ghost term
 associated with the broken symmetry $SU(N)/U(1)^{N-1}$
\beq
S_{\rm g.f.} + S_{\rm F.P.} = 
 -\frac{1}{2g^2} Tr ([X_\mu, A^\mu]^2)
 - \frac{1}{g^2}Tr ([X_\mu,b][A^\mu,c]),
 \eeq
the action  can be expanded   up to the second order of the
off-diagonal components $\tilde{A_{\mu}}, \tilde{\psi}$ as
\begin{eqnarray}
S_2+S_{\rm g.f.}
&=&\frac{1}{2g^2} \sum_{i\ne j}
\bigl( (x_\nu^i-x_\nu^j)^2 {{\tilde A}^{ij}_\mu}{}^*
{{\tilde A}^{ij}}{}^\mu
- \bar{\tilde \psi}^{ji}\Gamma^\mu (x^i_\mu-x^j_\mu) \tilde \psi^{ij}\n
&&+ (\bar{\xi}^i-\bar{\xi}^j) \Gamma^\mu \tilde \psi^{ij}
{\tilde{A_\mu^{ij}}}^*
+\bar{\tilde \psi}^{ji} \Gamma^\mu ({\xi}^i-{\xi}^j){\tilde{A_\mu^{ij}}}
\bigr).
\label{eq:S2comp}
\end{eqnarray}
The first and the second terms are the kinetic
terms for $\tilde{A}$ and $\tilde{\psi}$ respectively,
while the last two terms are
$\tilde{A}\tilde{\psi}\xi$ vertices. 
A  bosonic off-diagonal component $\tilde{A}_{\mu}^{ij}$
is transmuted to a fermionic
off-diagonal component $\tilde{\psi}^{ij}$
emitting a fermion zeromode $\xi^i$ or $\xi^j$.
This vertex conserves
$SU(N)$ indices $i$ and $j$.
Note that the propagators for $\tilde{A}$ and $\tilde{\psi}$
damp as $1/(x^i-x^j)^2$ or $1/(x^i-x^j)$ 
respectively.
\par
Integration over all the off-diagonal components gives an
 effective action for the zeromodes, $x_{\mu}^i$ and $\xi^i$:
\begin{eqnarray}
 \int d \tilde A d\tilde \psi d b dc \;
e^{ - (S_2 +S_{\rm g.f.}+S_{\rm F.P.})}
&=& \prod_{i<j} {\rm det}_{\mu\nu}
\bigl( \eta^{\mu\nu}+S^{\mu\nu}_{(ij)}\bigr)^{-1} \nonumber\\
&\equiv&
e^{-S_{\rm eff}^{\rm 1-loop} [X, \xi]} ,
\end{eqnarray}
where
\begin{equation}
S^{\mu\nu}_{(ij)}=\bar{\xi^{ij}} \Gamma^{\mu\alpha\nu}
\xi^{ij} \frac{x^{ij}_\alpha}{(x^{ij}) ^4}.  \label{eq:defS}
\end{equation}
Here $\xi^{ij}$ and $x_{\mu}^{ij}$ are abbreviations  for $\xi^i - \xi^j$
and $x_{\mu}^i- x_{\mu}^j $.
\par
The effective action can be expanded as
\begin{eqnarray}
S_{\rm eff}^{\rm 1-loop} [X, \xi]
&=& \sum_{i<j} tr \ln(\eta^{\mu\nu}+S^{\mu\nu}_{(ij)}) \nonumber \\
&=& - \sum_{i<j} tr
\bigl(\frac{S^4_{(ij)}}{4}
+\frac{S^8_{(ij)}}{8} \bigr) ,
\end{eqnarray}
which is a sum of all pairs $(ij)$ of space-time points.
Here the symbol $tr$ in the lower case stands for the trace for Lorentz
indices. Other terms in the expansion vanish due to the properties of 
Majorana-Weyl fermions in ten dimensions. Note that, since
 $S^{\mu\nu}_{(ij)}$ contains two fermion zeromodes, the first term
$S^4_{(ij)}$ contains 8 fermion zeromodes $\xi^{ij}$ along a link $(ij)$
and the second term $S^8_{(ij)}$ contains full 16 fermion zeromodes.
Hence, when integrating the fermion zeromodes,   the second term
acts as a delta function for the grassman variables 
$\delta^{16}(\xi_{i}-\xi_j)$. 
Integration over the fermion zeromodes 
gives  the final effective action 
$S_{\rm eff}^{\rm 1-loop} [X]$
for the bosonic eigenvalues 
$x_{\mu}^i$. 
\par
Several comments are in order.
The first comment is about supersymmetry. The one-loop effective action
$S_{\rm eff}^{\rm 1-loop} [X, \xi]$ has ${\cal N}=2$ supersymmetry
which is
a remnant of the original one:
\beqa
&&\cases{
\delta  x^i_\mu &  $= i \bar{\epsilon_1}\Gamma_\mu \xi^i $\cr
\delta  \xi ^i&  $=\epsilon_2$.
} \label{eq:susya}
\eeqa
Transformations for $\epsilon_1=\epsilon_2$ and 
$\epsilon_1=-\epsilon_2$ correspond to those  generated by
${\cal N}=1$ supersymmetry generator $Q$ and its covariant derivative $D$.
In this sense zeromodes of  $x_i$ and $\xi_i$ can be viewed as
 supercoordinates of ${\cal N}=1$
superspace. 
\par
The next  comment is infrared convergence. The partition function
is obtained by integrating over all the boson and fermion zeromodes.
A naive expectation is that, since the  bosonic coordinates 
originally parameterize the supersymmetric moduli, 
 the integral over $x_i$ diverges. 
However, as proved in \cite{AIKKT}, the integral converges for finite
$N$ to all orders of perturbation theory due to the damping property
of the propagators.
At one loop this infrared convergence can be easily seen since two
fermion zeromodes in 
$S^{\mu\nu}_{(ij)}$ always appear with a damping factor 
$1/(x^i-x^j)^3$ and a condition  for 
saturating  16 fermion zeromodes requires 
$1/r^{24}$ damping when two clusters of eigenvalues are put
apart with distance $r$.
 This shows that all points 
are gathered as a single bunch and hence space-time is inseparable.
This is consistent with the explicit calculations of the
partition function \cite{MNS}
\par
The one-loop effective action for $x_i$ is given by further integrating out
the fermion zeromodes  $\xi_i$. 
Non-vanishing contributions  in the expansion of 
$\exp(-S_{\rm eff}^{\rm 1-loop}[X,\xi])$ 
come from terms which 
saturate  all the fermion zeromode integrals  $d\xi_i$.
Since the terms $tr(S^4_{(ij)})$ or $tr(S^8_{(ij)})$ in the action
$S_{\rm eff}^{\rm 1-loop}$ contains 8 or 16 fermion zeromodes $\xi_{ij}$
respectively along a link $(ij)$, 
this expansion can be visualized as summing over  graphs
in which some of the ${}_N C_2$ pairs  among the 
$N$ space-time points $x_i$ ( $i=1...N$)  are connected by 
a term with 8 fermion zeromodes $tr(S^4_{(ij)})/4$ or a term
with 16 fermion zeromodes  
$(tr (S^4_{(ij)}))^2/32+tr(S^8_{(ij)})/8$:
\beqa
\int dXd\xi e^{-S^{\rm 1-loop}_{\rm eff}[x,\xi]}
&=& \int dXd\xi \sum_{G:{\rm graph}}\;\; \prod_{(ij):{\rm bond\; of \;} G}\n
&&               [(\frac{tr(S_{(ij)}^4)}{4})\;\; {\rm or}\;\;
                (\frac{1}{2}(\frac{tr(S_{(ij)}^4)}{4})^2
                +\frac{tr(S_{(ij)}^8)}{8})] \n
&=& \sum_{G:{\rm graph}} \int dX \ W[X;G].
\label{eq:graph}
\eeqa
Here $W[X;G]$ is a Boltzmann weight for a graph $G$ and configuration
${X}$.
In this way,  we arrive at network picture of the space-time. The
network is induced by fermion zeromode integrations.
\par
Integrations over the fermion zeromodes
give  interactions along a link $(ij)$ of the graph, which is
of order $1/(x^{ij})^{3}$ for
each fermion zeromode  between two points $(ij)$. That is, the terms
with 8 or 16 
fermion zeromodes give bond interactions of order 
$1/(x^{ij})^{12}$ or
$1/(x^{ij})^{24}$.\footnote{
Note that 
the bond interactions induced by terms with 8 fermion
zeromodes are not scalars and depend on  relative angles
between several links $x^i-x^j$ in a graph. On the other hand, 
interactions induced by terms with 16 zeromodes are scalars
since $S^8_{(ij)}$ acts as a delta function for $\xi_{ij}$.
}
In this manner  we obtain networks of space-time
coordinates connected by  bonds  with these interactions.
Since the bond interactions  suppress  connection  
between two distant  points at least by a factor of order $1/(x^{ij})^{12}$,
only closer points tend to be  connected and the network
becomes local. 
Also to saturate the grassman integral, we need $16(N-1)$
fermion zeromodes. Hence  number of bonds is of order $N$ which is much
smaller than possible number of pairs $N(N-1)/2$ 
in the large $N$ limit.
In this sense $N$ discretized space-time points are weakly bound.
All the allowed 
 networks must be summed in determining the dynamics of the eigenvalues.
 This reminds us of  summation of
all triangulations in the dynamical triangulation 
approach to quantum gravity.
We come back to this analogy in section 4..
\par
As shown in \cite{AIKKT}, if we neglect the first type of bond
interactions 
with  8 fermion zeromodes in (\ref{eq:graph}), 
the $\xi$ integration can be performed exactly
and the dynamics of eigenvalues is governed by the statistical
system  of a branched polymer. Since the Hausdorff dimension of 
the branched polymer is four,  we may conjecture that other interactions
will make  eigenvalue distribution a smooth four dimensional 
manifold, which may be classified as 
 a new universality class near the branched 
polymer phase.
Numerical simulation is still under investigation.
\section{Local Gauge Invariance}
\setcounter{equation}{0}
Once we describe the  space-time as 
dynamically generated distribution of the eigenvalues,  low energy effective
theory in the space-time
 can be obtained by solving  dynamics of   fluctuations around the
background $X_{\mu}$. 
Both of  the space-time $X_{\mu}$ and matters $\tilde{A}_{\mu}$
 are unified in the same matrices $A_{\mu}$ 
and should be determined dynamically.
Low energy fluctuations 
 are in general composites of $A_{\mu}$ and $\psi$, and
it is natural from the analysis of the 
Schwinger-Dyson equation for the  Wilson loops
that a local operator in the  space-time 
is given by a microscopic limit of the Wilson loop operators,
such as 
\beq
w(k;O) = Tr [ O(A,\psi ) \ exp (i k^{\mu}A_{\mu}) ].
\label{localop}
\eeq
Here $O(A, \psi )$ is some operator made of $A_{\mu}$ and $\psi$. 
In order to identify the total momentum of this operator as
$k$, the operator $O(A, \psi )$ should be invariant under  a constant
shift of $A_{\mu}$, that is,  translation in the 
space-time coordinates.
\par
In the first approximation around the diagonal background $X_{\mu}$,
the coordinate representation of this  operator is given by  
\beqa
\hat{w}(x;O) &=& \int  {d^{10}k \over (2 \pi)^{10}} \ 
exp(- i  k^{\mu} x_{\mu}) \ w(k;O)
\n
&\sim& \sum_{i=1}^{N} O_{ii} \delta^{(10)}({\bf x}-{\bf x_i}).
\eeqa
Here we have replaced $A_{\mu}$ by $X_{\mu}+\tilde{A}_{\mu}$ and take 
the leading term. $O_{ii}$ is the $ii$ component of the operator $O$.
Due to the delta function,  
the operator has a support   only on the area where
eigenvalues distribute.
 Vanishing of the operator $\hat{w}(x;O)$ 
outside of  the area of the distributed  eigenvalues
 supports our interpretation of the space-time.
Of course if we take into account higher terms in the expansion, 
the operator becomes dim and extended around the background $X_{\mu}$.
\par
We can apply a similar analysis to strings which propagate in the space-time.
In the $1/N$ expansion,
correlation  between Wilson loop operators can be evaluated by
summing over all surfaces made of Feynman diagrams 
connecting the Wilson loops at the boundary.
This surface is interpreted as  string world sheet connecting 
strings at the boundaries.
Each $SU(N)$ index $x_{\mu}^i$ of  a loop in  the diagrams represents  a
coordinate on the world sheet  and 
it takes value in the eigenvalue distribution in the leading
approximation around the diagonal background $X_{\mu}$. 
Hence string world sheet evolves only in the  space-time of the  
eigenvalue distribution and again supports our interpretation of 
the space-time.
\par
It is generally difficult  to 
obtain how fluctuations  propagate in the eigenvalue distribution,
which  is reminiscent of the QCD effective theory: 
Excitations are expressed as composite operators of microscopic variables
 and their low energy dynamics can be discussed  only 
through  the  symmetry argument, 
namely the argument based on the chiral symmetry.
Also in our case  we will show that there are eigenvalue distributions
around which  symmetry arguments  allow us to discuss    low energy 
dynamics for some excitations. 
Suppose that 
eigenvalue distribution  forms clusters consisting  of $n$ eigenvalues.
At length scale much larger than the size of each cluster, 
the  $SU(N)$ symmetry is
broken down to $SU(n)^{m}$ where $m=N/n$.
We can 
expand $A_{\mu}$ and $\psi$ around such a background $X_{\mu}$
similarly to the analysis in the previous section.
First write $A_{\mu}$ and $\psi$ in  block forms:
\beqa
A_{\mu}&=& \left( \begin{array}{llll}
                 A_{\mu}^{11} &A_{\mu}^{12} & ... &  \\
                 A_{\mu}^{21} &A_{\mu}^{22} & & \\
                  \vdots            &             & \ddots &   \\
                  &&& A_{\mu}^{mm}
         \end{array} \right), \n
\psi &=& \left( \begin{array}{llll}
                 \psi^{11} &\psi^{12}& ...& \\
                 \psi^{21}& \psi^{22}&& \\
                 \vdots & & \ddots &   \\
                  &&& \psi^{mm}
         \end{array} \right).
\eeqa
Each block $A_{\mu}^{ij}$ or $\psi^{ij}$ 
is an $n \times n$ matrix and  the diagonal blocks
can be further decomposed
\beqa
A_{\mu}^{ii} &=& x_{\mu}^{i} {\bf 1}+\tilde{A}_{\mu}^{ii} \n
\psi^{ii} &=& \xi^i {\bf 1}+\tilde{\psi}^{ii}
\eeqa
where $ {\bf 1}$ is an $n \times n$ unit matrix and
$tr \tilde{A}_{\mu}^{ii}=0$. Here $tr$ means trace for the
submatrix of $n \times n$.
We interpret each cluster of the eigenvalues 
as a space-time point with an internal 
structure $SU(n)$. 
Since each $SU(n)$ symmetry acts on the variables at 
the position $i$ independently,
the unbroken  $SU(n)^m$ symmetry can be regarded as  local gauge
symmetry. 
Indeed under a gauge
transformation $g$ of the unbroken $SU(n)^m$ symmetry
\beqa
g=\left( \begin{array}{llll}
                 g_1 &    &  & \\
                     &g_2 &  & \\
                     &    & \ddots &   \\
                     &    &       & g_m \\
         \end{array} \right)  \in SU(n)^m \subset   SU(N),
\eeqa
the diagonal block fields, $\tilde{A}_{\mu}^{ii}$ and $\tilde{\psi}^{ii}$,
transform as adjoint matters
(i.e. site variables in  the lattice gauge theory)
while the off-diagonal block fields,   $A_{\mu}^{ij}$ and $\psi^{ij}$,
as gauge connections
(i.e. link variables):
\beqa
\tilde{A}_{\mu}^{ii} \rightarrow g_i \tilde{A}_{\mu}^{ii} g_i^{\dagger}
\n
\tilde{\psi}^{ii} \rightarrow g_i \tilde{\psi}^{ii} g_i^{\dagger}
\n
A_{\mu}^{ij} \rightarrow g_i A_{\mu}^{ij} g_j^{\dagger}
\n
\psi^{ij} \rightarrow g_i \psi^{ij} g_j^{\dagger}.
\eeqa
Some of the dynamics for
low energy excitations is governed by this
local gauge invariance.  Gauge fields live on the links   and transform
as the link variables in  the lattice gauge theory. In our case, we have
too many such fields (at least  10 boson fields $A_{\mu}^{ij}$
for a link $(ij)$ ) but only one unitary link variable
 is  assured to be
massless by the gauge symmetry and others will acquire mass dynamically.
Therefore, in deriving low energy effective theory, 
 we first apply polar decomposition to $A_{\mu}^{ij}$ 
into unitary and hermitian degrees of freedom and 
identify all the  unitary components 
in various off-diagonal block fields  
by setting  them one common field 
 $U^{ij}$ on each link. 
\par
Among various fields 
in the dynamically generated random lattice, 
some fields will  survive in the low energy effective theory.
Here let us consider simple fields made of a single microscopic
variable $A_{\mu}$ or $\psi$. As we will see soon, 
the off-diagonal block fields  become massive except the
unitary component $U^{ij}$.
The gauge field $U^{ij}$ 
is assured to be massless due to the gauge invariance. 
Next, stability of the cluster type eigenvalue distribution we have assumed
requires that  the diagonal bosonic fields $A_{\mu}^{ii}$, which break
this distribution, should become massive dynamically.
Finally, the diagonal fermionic block field $\psi^{ii}$ 
can be massless and we should investigate their low energy action.
\par
We henceforth integrate out 
the off-diagonal blocks first while keeping the unitary
degrees of freedom $U^{ij}$ and obtain an effective action for
other variables which  survive in the low energy effective theory.
\par
We add a gauge fixing term for broken generators $SU(N)/SU(n)^m$ 
\beq
S_{\rm g.f.}= \frac{1}{2g^2} \sum_{i \ne j}
 tr|x_{\mu}^{ij} A_\mu^{ij}|^2,
\eeq
and also its associated Fadeev-Popov term.
The action can be expanded as before. Lower order  terms in the action
are composed of the following terms:
\begin{eqnarray}
S_2+S_{\rm g.f.}
&=& \frac{1}{2g^2} \sum_{i\ne j}
tr \bigl( (x_\nu^{ij})^2 |A^{ij}_\mu|^2
- \bar{\psi}^{ji}\Gamma^\mu x^{ij}_\mu  \psi^{ij}\n
&& -2 \bar{\psi}^{ii}\Gamma^\mu (A_\mu^{ij} \psi^{ji}-\psi^{ij}A_\mu^{ji})
\bigr).
\label{lowerorder}
\end{eqnarray}
The first and the second terms are the kinetic terms for the off-diagonal
block fields $A_{\mu}^{ij}$ and $\psi^{ij}$.
The third term corresponds to the term of the zeromode $\xi$ insertion
in the previous section.
Integrating out the
fermion fields in the off-diagonal blocks $\psi^{ij}$, 
the third term in (\ref{lowerorder}) becomes
\beq
 {1 \over g^2}\sum_{i\ne j}tr(\bar{\tilde \psi}^{ii}
\Gamma^{\mu \lambda \nu} A_\mu^{ij}{x^{ij}_{\lambda}  \over(x^{ji})^2}
 (A_\nu^{ji} \tilde{\psi}^{ii}-\tilde{\psi}^{jj} A_\nu^{ji})).
\label{Sint}
\eeq
It consists  of  a mass-like term proportional to 
$\bar{\tilde{\psi}}^{ii}\Gamma^{\mu \lambda \nu} \tilde{\psi}^{ii}$
and a hopping term $\bar{\tilde{\psi}}^{ii} \tilde{\psi}^{jj}$
between $i$ and $j$.
Integration over  $A_{\mu}^{ij}$
can be performed, with its unitary component $U^{ij}$ kept fixed,
by replacing
\beqa
\overline{A_{\mu}^{ij}} &=& 0 \n
\overline{A_{\mu}^{ij} \otimes A_{\nu}^{ji}} &=& 
g^2 {\delta_{\mu \nu} \over (x^{ij})^2} U^{ij} \otimes U^{ji}.
\label{replacement}
\eeqa
In the second equation, the damping factor $1/(x^{ij})^2$ 
corresponds to the 
propagation of the hermitian degrees of freedom while the 
appearance of the link variable $U^{ij}$ corresponds to keeping the unitary
degrees of freedom. 
This replacing makes the mass-like term in 
 (\ref{Sint}) vanish and the hopping term 
\beq
- g^2 \sum_{i\ne j}tr(\bar{\tilde{\psi}}^{ii} \Gamma^{ \lambda}
{x^{ij}_{\lambda}  \over(x^{ij})^4}U^{ij} \tilde{\psi}^{jj} U^{ji}).
\label{adjointfermion}
\eeq
This is a gauge invariant kinetic term for an  adjoint
fermion  field   $\psi^{ii}$ on  a (random) lattice
generated dynamically by  the distributed eigenvalues.
The fields $\tilde\psi^{ii}$ can hop between any pair of points
in the  space-time, but since the hopping parameter is suppressed 
by $1/(x^{ij})^3$,  the propagation is expected to  become local  
  in the continuum limit. 
\par
Other terms in the action generate a plaquette action for the gauge
field $U^{ij}$ as follows. Relevant terms in the action are 
\beq
S_{4}= {1 \over g^2} \sum_{i \ne j \ne k \ne l} tr
(A_{\mu}^{ij} A_{\nu}^{jk} -A_{\nu}^{ij} A_{\mu}^{jk})
A_{\mu}^{kl} A_{\nu}^{li}.
\eeq
By integrating out the hermitian degrees of freedom of the
off-diagonal blocks  
with the procedure  (\ref{replacement}), this action
itself vanishes $\bar{S_4}=0$. However  interactions
generated by $(S_4)^2$ induce a  kinetic term for the gauge
field;
\beq
\overline{(S_4)^2} \sim  \sum_{i \ne j \ne k \ne l} 
{g^4 \over (x^{ij})^2(x^{jk})^2 (x^{kl})^2 (x^{li})^2  }
tr(U^{ij} U^{jk} U^{kl} U^{li}) tr(U^{il}U^{lk}U^{kj}U^{ji}).
\label{plaquette}
\eeq
This is the  plaquette action generated by a Wilson loop for
the adjoint representation and hence the  gauge field $U^{ij}$
indeed propagates in  the space-time of   eigenvalue distribution.
Again the gauge field can hop between any pair of space-time
points, but the hopping is suppressed by $1/x^8$ for distant points
and we will recover locality in the continuum limit.
\par
To summarize this section, 
supposing  that   distribution of the eigenvalues  consists of
small clusters with size $n$, we have shown that the low energy 
effective theory contains several massless fields such as 
the gauge field associated with the local $SU(n)$ gauge symmetry
and  fermion  field in the adjoint representation of $SU(n)$
gauge symmetry. Gauge invariant kinetic terms were also
derived. 
Thus our system  is a lattice gauge theory 
on a dynamically generated random lattice.
It is invariant under a permutation for the set of the $N$ discrete
space-time points, 
since the permutation group $S_{N}$ is a subgroup of 
the original $SU(N)$ symmetry. 
It is the most different point from the ordinary lattice gauge theory
on a fixed lattice,  
which   becomes important in deriving the 
diffeomorphism invariance of our model.
We will come back to this point in the next section.
Although the permutation invariance requires 
that all space-time points are equivalent, 
 locality in  the space-time will 
be assured due to the suppression of the hopping terms between  distant
points. In general, however, we need  a sufficient power for  the damping 
of the  hopping terms in order to assure locality in the continuum limit.
Though  we do not yet know the   real condition for locality,
we expect that  terms with lower  powers 
are canceled due to   supersymmetry or 
by averaging over gauge fields. 
\section{Gravity and Diffeomorphism Invariance}
\setcounter{equation}{0}
As we have seen in section 2, the one-loop effective action for 
the space-time points  is described as a statistical 
system of $N$  points whose coordinates are $x_{\mu}^{i}$.
Integration over the fermion zeromodes $\xi$ gives  
the Boltzmann weight, which depends on a graph (or network) connecting
the space-time  points locally by order $N$ number of bond interactions;
 \begin{equation}
Z= \sum_{G:{\rm graph}} \int dX \ W[X;G].
 \end{equation}
$ W[X;G]$ is a complicated function of  a configuration $X$ and a 
graph $G$. An important property is that the weight is suppressed
at least by a damping  $1/(x^i-x^j)^{12} $ 
when two  points $i$  and $j$
are connected.
This system is, of course,  invariant under permutations $S_N$ 
\footnote{In this section we consider general eigenvalue distributions
in which all eigenvalues have indegenerate  space-time coordinates. 
If we take the cluster type distribution considered in the previous
section, the permutation symmetry responsible for the diffeomorphism 
invariance should be $S_{N/n}.$
}
of $N$ space-time
points, which is a subgroup of the original symmetry $SU(N)$,
while  the  Boltzmann weight for  each graph $G$ is not.
The invariance  is realized by summing  over all possible graphs.
In other words the system becomes permutation invariant by 
rearrangements of  bonds in the  network of space-time points.
This reminds us of  the dynamical triangulation approach to 
quantum gravity \cite{ambjorn},  
where the diffeomorphism invariance is believed to arise from 
summing all possible triangulations.
It is amusing that our system satisfies both of locality
and permutation invariance simultaneously 
by summing  over  all possible graphs.
\par
In this section we  see that the permutation
invariance of our system actually leads to the diffeomorphism invariance.
To see how  the background metric  is encoded
in the  effective action for low energy excitations, 
let us consider, as an example,  a  scalar field $\phi^i$ propagating in 
distributed eigenvalues.  The effective action will be given by
\begin{equation}
S= \sum_{i,j} {(\phi^i-\phi^j)^2 \over 2} f(x^i-x^j) + \sum_{i} m (\phi^i)^2
\end{equation}
where $f(x)$ is a function  decreasing  sufficiently fast at infinity
to assure
locality in  the space-time. Introducing  the density function of the 
eigenvalues
\begin{equation}
\rho(x)= \sum_i \delta^{(10)}(x-x^i)
\end{equation}
and a continuous  field $\phi(x)$ which  satisfies 
 $\phi(x^i)=\phi^i$,
the action can be rewritten as
\begin{equation}
S = \int dx dy \langle \rho(x) \rho(y) \rangle 
{(\phi(x)-\phi(y))^2 \over 2} f(x-y)
+ m \int dx  \langle \rho(x) \rangle \phi(x)^2.
\end{equation}
Here  the expectation $\langle ...\rangle$
for the density and the density  correlation  means that we have
taken average over  configurations $X$ and   networks  $G$ of 
the space-time points. 
Normalizing  the density correlation in terms of the density
\begin{equation}
\langle \rho(x) \rho(y) \rangle = \langle \rho(x) \rangle \langle \rho(y) \rangle
(1+c(x,y))
\end{equation}
and expanding $\phi(x)-\phi(y) = (x-y)_{\mu}\partial^{\mu} \phi(x) +
\cdot \cdot \cdot$, the action becomes
\begin{eqnarray}
S &=& {1 \over 2} \int dx \langle \rho(x) \rangle  \left[  \int dy 
\langle \rho(y) \rangle (x-y)_{\mu}(x-y)_{\nu} f(x-y) (1+c(x,y)) \right] \n &&
\partial^{\mu} \phi(x) \partial^{\nu} \phi(x)
 + m \int dx \langle \rho(x) \rangle \phi(x)^2 \cdot \cdot \cdot.
\label{phiaction}
\end{eqnarray}
This expansion shows  that  the field $\phi(x)$ propagating 
in  the eigenvalue distribution 
feels  the density correlation as the  background metric
while  the density itself as  vacuum expectation value of 
the dilaton field. Namely we can identify
\begin{eqnarray}
g_{\mu \nu}(x) &\sim&  \int dy 
\langle \rho(y) \rangle (x-y)_{\mu}(x-y)_{\nu} f(x-y) (1+c(x,y)) 
\label{metric} \\
\sqrt{g} e^{- \Phi(x)} &\sim& \langle \rho(x) \rangle.
\label{dilaton}
\end{eqnarray}
If the density correlation respects the original translational and 
rotational symmetry, that is, if they are not spontaneously broken,
the metric becomes flat $g_{\mu \nu} \sim \eta_{\mu \nu}$.
(Normalization can be absorbed by the dilaton vev.)
The fact that the background metric is  encoded in the density
correlations as above indicates that our system is general covariant
even though the IIB matrix model action (\ref{action}) defined
in flat ten dimensions does not have  manifest general covariance.
\par
Then let us see how the diffeomorphism invariance  is realized in our 
model.
The action  (\ref{action})
is invariant under the permutation $S_N$ of the eigenvalues, which is a
subgroup of $SU(N)$.
Under a permutation 
\begin{equation}
x^i \rightarrow x^{\sigma(i)} \ \ \mbox{for} \ \ \sigma \in S_N,
\label{perm}
\end{equation}
the field $\phi^i$ transforms into $\phi^{\sigma(i)}$.
Then, from the definition of the continuous  field $\phi(x)$, we should 
extend the transformation (\ref{perm}) into $x$,
\begin{equation}
x \rightarrow \xi(x)
\label{permx}
\end{equation}
such that $\xi(x^i) = x^{\sigma(i)}$.
Under this transformation, the eigenvalue density transforms as 
a scalar density and the  field $\phi(x)$ as a scalar field.
On the other hand, the metric transforms as a second rank tensor, if the 
function $f(x-y)$ decreases rapidly around $x=y$
and the $y$ integral in (\ref{metric})
has support  only near $y=x$. 
The tensor property of the metric
 is also required  from the invariance of the 
action under the  transformation (\ref{permx}).
In this way,  the 
invariance under the permutation of the eigenvalues leads to 
the invariance of  the low energy effective action 
under general coordinate transformations.
\par
Background metric is encoded in the  density correlation of the
eigenvalues. Since we have started  from the Poincare invariant
type IIB matrix model action (\ref{action}), the density correlation is 
expected to be translational and rotational invariant
and we may  obtain low energy effective action  in a flat background.
A nontrivial background can be induced dynamically 
if Lorentz symmetry is spontaneously broken and
the  eigenvalues are non-trivially distributed.  
\par
A nontrivial background can be also described  by condensing a graviton
operator \cite{yoneya-nishi}. 
Bosonic parts of graviton and dilaton operators are given by
\begin{eqnarray}
S_{\mu \nu}(k) &\sim&   Tr(F_{\mu \lambda} {F^{\lambda}}_{\nu} e^{ik \cdot A})
+ (\mu \leftrightarrow \nu)  \label{graviton}
\\ 
D(k) &\sim&   Tr(F^2 e^{ik \cdot A}). \label{dilaton}
\end{eqnarray}
Condensation of them induces extra terms in IIB matrix model action;
\begin{equation}
S_{\mbox{cond}} = \int dk (\sum h^{\mu \nu}(k) S_{\mu \nu}(k) + h(k) D(k)).
\end{equation}
We can similarly obtain an effective action for fluctuations  around a diagonal
background from this modified matrix model action.
 Condensation of dilaton changes 
the Yang-Mills coupling constant $g$ locally in the space-time.
Since $g$ is the only dimensionful constant in our model and thus determines
the fundamental length scale, a local change in $g$ will lead to 
a local change in the eigenvalue density.
This is consistent with our earlier discussion that  the dilaton
expectation value is encoded in the  eigenvalue density. 
On the other hand, condensation of graviton induces asymmetry of
the space-time. For condensation of $k=0$ graviton mode, it is 
obvious that the condensation can be compensated by a field
redefinition of matrices $A_{\mu}$
\beq
A_{\mu} \rightarrow (\delta_{\mu}^{ \nu} + h_{\mu}^{ \nu}) A_{\nu},
\eeq
and the two models, the original IIB matrix model and the modified
one with the $k=0$ graviton condensation, 
are directly related through the above field redefinition.
The density of the eigenvalues is mapped accordingly and the density
correlation is expected to become asymmetric in the
modified matrix model.
For more general condensation,
if  the  graviton operator $\hat{S}_{\mu\nu}(x)$ 
(coordinate representation of  (\ref{graviton}))
 changes only  local property of dynamics of the eigenvalues,
 the  density correlation will become asymmetric locally in the space-time
around  $x$
and therefore induces a local change in the background metric.
\par
Our low energy effective action is formulated as a lattice gauge theory
on a dynamically generated random lattice. Since  the lattice itself is 
generated dynamically from  matrices, we must sum over all possible
graphs.  In this way,  our system 
is permutation $S_N$ invariant,  which is responsible for the
diffeomorphism invariance. 
The background metric is encoded in the density correlation of the 
eigenvalues and the low energy effective action becomes manifestly
general covariant.
The graviton operator is represented as fluctuation around the
background space-time and constructed from the off-diagonal components of the 
matrices. Microscopic derivation of the 
propagation of the graviton is difficult to obtain,  
but  once we have clarified the
underlying diffeomorphism symmetry,
it is natural that the low energy effective action for the graviton
is described by the Einstein Hilbert action. 
Employing this diffeomorphism invariance and the supersymmetry,
we will be able to  derive the low energy behavior of the graviton multiplets,
which will be reported in a separate paper. 
\section{Conclusions and Discussions}
In this paper we have discussed a possibility to interpret the
space-time in type IIB matrix model as distributed  eigenvalues.
We have shown that, if we suppose  that   eigenvalue
distribution consists of small clusters with size $n$, 
the low energy theory acquires $SU(N)$ local space-time gauge symmetry.
This gauge invariance can  predict the existence of 
the gauge field  propagating
in the  space-time of  distributed eigenvalues.
Also we have obtained a gauge invariant kinetic action for a
fermion in the adjoint representation of $SU(N)$. 
Low energy behavior  for these fields is described
in terms of  a lattice gauge theory on a dynamically generated random lattice
and hence supports our interpretation of the space-time.
\par
Since the type IIB matrix model is proposed as a constructive
definition of superstring, we need to show the
existence of the massless graviton and the diffeomorphism invariance
of the low energy effective theory. One plausible argument for the
existence of   massless graviton 
is  based on the maximal supersymmetry ($D=10$ ${\cal N}=2$ susy).
We have shown that
the diffeomorphism invariance of our model is originated in 
the invariance under  permutations  of the eigenvalues.
Our model realizes the invariance in an interesting way by 
summing all possible graphs connecting the space-time points.
The diffeomorphism invariance  restricts  the  low energy behavior of the model
and gives another reasoning for the existence of massless graviton.
Background metric for propagating  fields is encoded in the density
correlation of the  eigenvalues while the dilaton vev is encoded in the 
 eigenvalue density. 
Curved background can be described as a nontrivial distribution 
of eigenvalues whose density correlation behaves inhomogeneously.
\par
Both of these fundamental symmetries, local gauge symmetry and the
diffeomorphism symmetry, are originated in the $SU(N)$ invariance of the 
matrix model. 
These symmetries  have
been attempted to unify by many physicists, beginning with 
Kaluza and Klein \cite{Kaluza}.
It is  amusing that our matrix model can unify them in a natural way
 and  complicated structures are  hidden in such a simple matrix model.
\par
There are still several problems unanswered.
One   is the low energy supersymmetry in the  effective theory
in  a dynamically generated distribution of eigenvalues.
Since we have obtained a massless gauge boson and an adjoint fermion,
we may expect ${\cal N}=1$ low energy supersymmetry.
We need to clarify how ${\cal N}=2$ supersymmetry in type IIB matrix
model acts on the low energy fields.
Since ${\cal N}=1$ ten dimensional Yang-Mills theory is anomalous
by itself, other fields (e.g. supergravity multiples) must  be required
massless to cancel anomalies. Also the low energy  gauge group might be
restricted or projected.
\par
Next problem is locality in the low energy effective theory.
The gauge invariant kinetic terms (\ref{adjointfermion})
 or (\ref{plaquette}) have  hopping parameters suppressed due to the
inverse powers of distances, but the suppression  seems insufficient to assure
locality or a well behaved infrared property. 
As for the action  (\ref{adjointfermion}),  we may 
obtain more suppressed hopping terms since the link variable 
$U^{ij}$ between  distant  points can fluctuate much and 
averaging over the gauge field may reduce the strength of
 effective hopping parameters.
\par
We have also more fundamental or conceptual issues.
We have obtained nonabelian gauge symmetry from type IIB matrix model
by assuming a particular eigenvalue distribution. 
This indicates that this  vacuum  is not a perturbative vacuum 
of type IIB superstring. Instead we may wonder if this is
 a perturbative vacuum of heterotic string or type I string realized 
in a nonperturbative way within type IIB matrix model.
As we saw in the introduction,  our matrix model 
contains both of the world sheets of the 
fundamental  IIB string and the D-strings.
By a semiclassical correspondence (\ref{correspondence}),
we have identified 
 IIB superstring in the Schild gauge where
 $tr$ is interpreted  as  integration over a D-string world sheet. 
We can also construct an  F-string world sheet 
in terms of surfaces made of 
Feynman diagrams whose $SU(N)$ index  represents
a space-time coordinate of a world sheet point.
In both cases, if we assume an eigenvalue distribution consisting of
small clusters, an internal structure appears on the world sheet
and hence current algebra may arise.
\par
Another issue is how to describe global topology in type IIB matrix
model. The simplest example is a torus compactification.
A  possible procedure of a torus compactification \cite{Taylor}
is  to identify $A_{\mu}$ with $A_{\mu} + R_{\mu}$
by embedding a derivative operator into our matrix configuration.
Therefore  $N$ is taken infinity from the beginning.
Since this procedure has a subtlety in the large $N$ limit,
we need a careful examination of the double scaling limit.
\par
We also do not yet know how we can describe chiral fermions 
in lower dimensions after compactification. Our compactification
procedure looks  different from the  ordinary Kaluza-Klein 
compactification  but there is some 
similarity. In the  ordinary case,
isometry of a compactified space leads to local gauge symmetry
in four dimensional space-time while in our case 
the internal structure arises from 
assuming the cluster type eigenvalue distribution.
In our description of four dimensional space-time,  eigenvalues
are 
confined in   four directions and shrunk in the other six
directions.  Our compactification therefore can be regarded as a segment
compactification in an ordinary picture.
We need to investigate how chiral fermions can arise in such a
compactified space-time.
\par
These problems and issues are under investigations and we want to discuss more
in  the near future.
\section*{Acknowledgments}
We would like to thank  Y. Kitazawa, T. Tada and A. Tsuchiya 
for discussions and especially H. Aoki
for the collaboration in the early stage. 

\end{document}